\documentclass[a4paper,12pt]{iopart}
\usepackage{iopams}
\usepackage[dvips]{epsfig}
\usepackage[dvips]{graphicx}
\usepackage{psboxit}

\begin{document}

\title[Response of the Brazilian GW detector to signals from a BH ringdown]{Response of the Brazilian gravitational wave
detector to signals from a black hole ringdown}

\author{C\'esar A. Costa and Odylio D. Aguiar}
\ead{cesar@das.inpe.br and odylio@das.inpe.br}
\address{Instituto
Nacional de Pesquisas Espaciais - Divis\~ao de Astrof\'isica\\Av.
dos Astronautas,1758 - Jd. Granja - CEP 12227-010\\S\~ao Jos\'e
dos Campos - SP - Brazil }
\author{Nadja S. Magalh\~aes}
\ead{nadjam@ita.br}
\address{Instituto Tecnol\'{o}gico de Aeron\'{a}utica -
Departamento de F\'{i}sica\\ Pra\c{c}a Marechal Eduardo Gomes, 50
- Vila das Ac\'acias - CEP 12228-900\\S\~ao Jos\'e dos Campos - SP
- Brazil}

\begin{abstract}
It is assumed that a black hole can be disturbed in such a way
that a ringdown gravitational wave would be generated. This
ringdown waveform is well understood and is modelled as an
exponentially damped sinusoid. In this work we use this kind of
waveform to study the performance of the SCHENBERG gravitational
wave detector. This first realistic simulation will help us to
develop strategies for the signal analysis of this Brazilian
detector. We calculated the signal-to-noise ratio as a function of
frequency for the simulated signals and obtained results that show
that SCHENBERG is expected to be sensitive enough to detect this
kind of signal up to a distance of $\sim 20\mathrm{kpc}$.
\end{abstract}
\pacs{04.80.Nn,95.55.Ym}
\submitto{\CQG}

\section{Black hole ringdown}

Ringdown waveforms originate from a small perturbation of a
spinning black hole (BH) and can be modelled as an
exponential-damped sinusoid. The central frequency $f$ of the
fundamental quadrupolar mode and the quality factor $Q$ depend on
the mass $M$ and the spin ($S=\hat{a}GM^2/c$) of the BH. They can
be approximated by an analytic fit
\cite{echeverria1989,leaver1985} and are given by
\begin{equation}\label{eq:f0}
f\approx
32\mathrm{kHz}\left[1-0.63\left(1-\hat{a}\right)^{\frac{3}{10}}\right]
\left(\frac{M}{M_\odot}\right)^{-1}, ~~~\mathrm{and} ~~~Q\simeq
2\left(1-\hat{a}\right)^{-\frac{9}{20}}.
\end{equation}
\par The value of the dimensionless spin parameter $\hat{a}$ is
zero in the Schwarzschild limit (non-rotating BH) and one in the
extreme Kerr limit (maximum rotational speed), so $Q>2$.
\par The relative strain caused by any metric perturbation on the detector
depends on the position of the BH in the sky and the relative
orientation of its spin axis to the local zenith of the detector.
The averaged strain produced by the tensorial GW components can be
obtained by \emph{rms} averaging over all possible angles and the
result is $\langle h_{GW} \rangle_{\mathrm{angle}}=Aq(t)$, where
$q(t)$ is the damping function that depends on the rotation speed
and on the central frequency of the BH:
\begin{equation}\label{eq:qt}
q(t)=(2\pi)^{\frac{1}{2}}e^{-\pi f t/Q}\cos(2\pi f t)
\end{equation}
for $t\geq 0$ (we set the time origin on $t=0$). The amplitude $A$
also depends on the fraction $\epsilon$ of the total mass-energy
radiated and on the BH distance from the Earth ($r$), and it is
given by
\begin{equation}\label{eq:A}
A \simeq 2.415\times
10^{-21}Q^{-\frac{1}{2}}\left[1-0.63(1-\hat{a})^{\frac{3}{10}}
\right]^{-\frac{1}{2}}\left(\frac{\mathrm{Mpc}}{r}\right)\left(\frac{M}{M_\odot}\right)
\left(\frac{\epsilon}{0.01}\right)^{\frac{1}{2}}.
\end{equation}
\par The quality factor roughly gives the number of coherent cycles
for the waveform \cite{creighton1999}. It affects the damping
($1/Q$), which decreases with increasing rotational speed.  To
generate the signal that was introduced in the model of the
detector we chose one (among many) of the sets of parameters that
made the central frequency of the BH quadrupolar normal mode
coincident with one of SCHENBERG resonant quadrupolar frequencies
($3172.5.6\mathrm{Hz}$, $3206.3\mathrm{Hz}$ and
$3240\mathrm{Hz}$).
\begin{figure}\label{fig:parameters}
\begin{center}
\includegraphics[width=8.5cm]{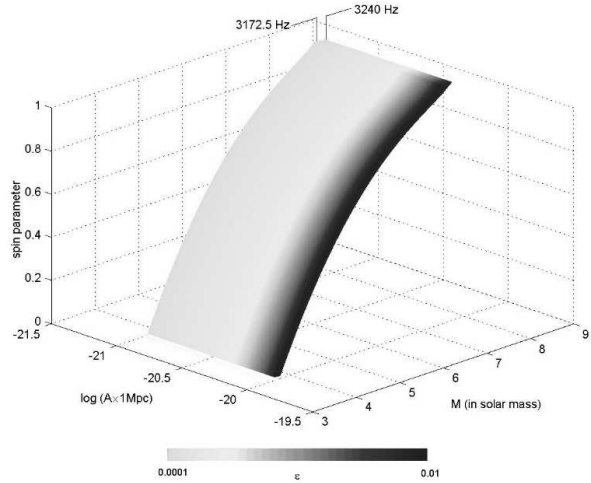}
\caption{Volume formed by sets of parameters that correspond to
frequencies varying from $f=3172.5\mathrm{Hz}$ to
$3240\mathrm{Hz}$.}
\end{center}
\end{figure}
We obtain a certain volume in the space of parameters shown in
figure \ref{fig:parameters} by fixing two boundary frequencies for
the BH quadrupolar mode and then making those parameters to vary.
Consequently any set of parameters inside that volume is within
SCHENBERG bandwidth.

\section{The simulated signal}\label{sec:signal}

We chose $\epsilon=0.01$, $M=4M_\odot$ and $\hat{a}=0.16$ to
generate the signal that we used to calculate the signal-to-noise
ratio. We believe that this efficiency is not overestimated since
similar values have been found in the literature
\cite{creighton1999,GRASP}.
\begin{figure}\label{fig:waveform}
\begin{center}
\begin{tabular}{c@{\qquad}c}
\mbox{\epsfig{file=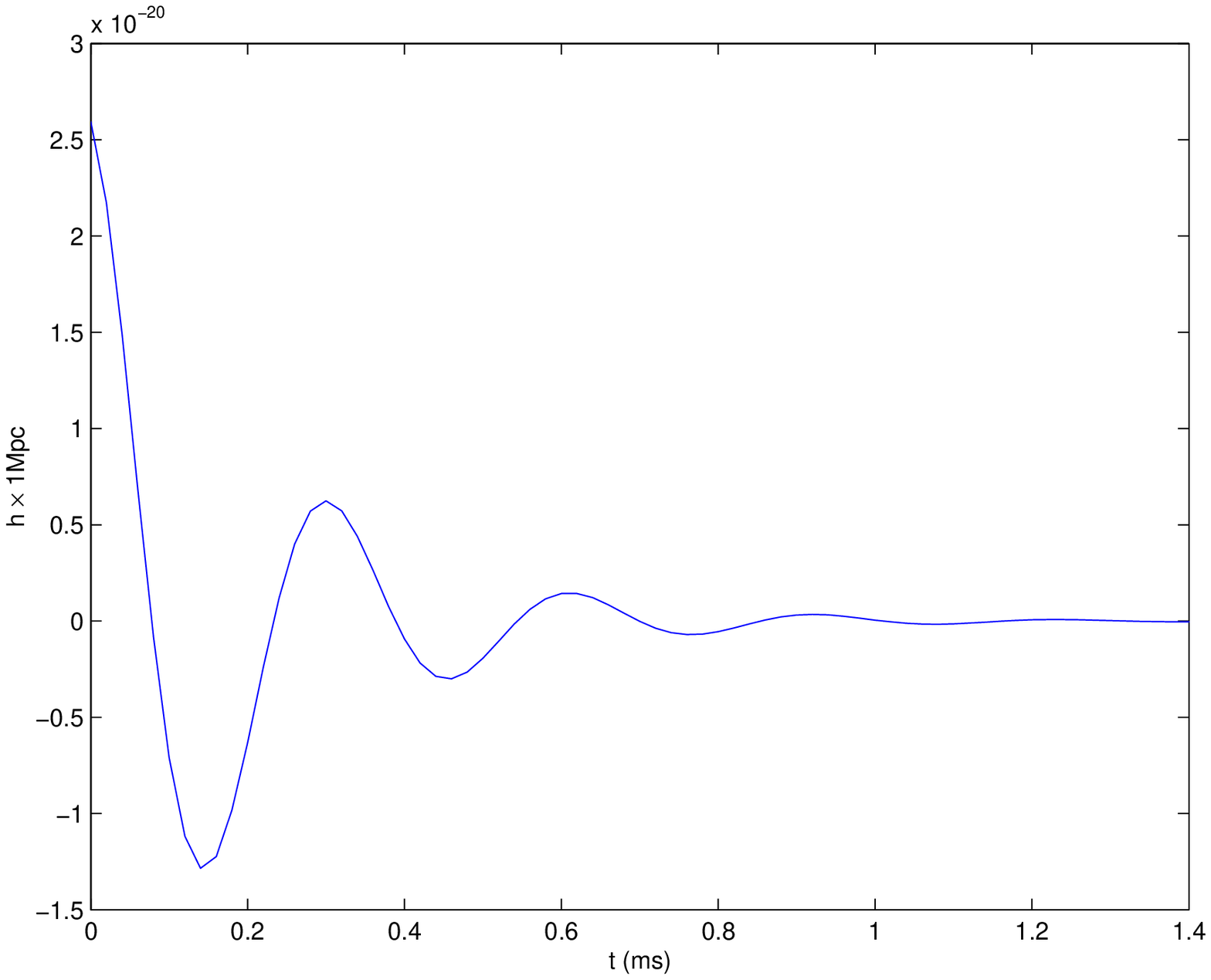, width=5cm, angle=0}}   &
\mbox{\epsfig{file=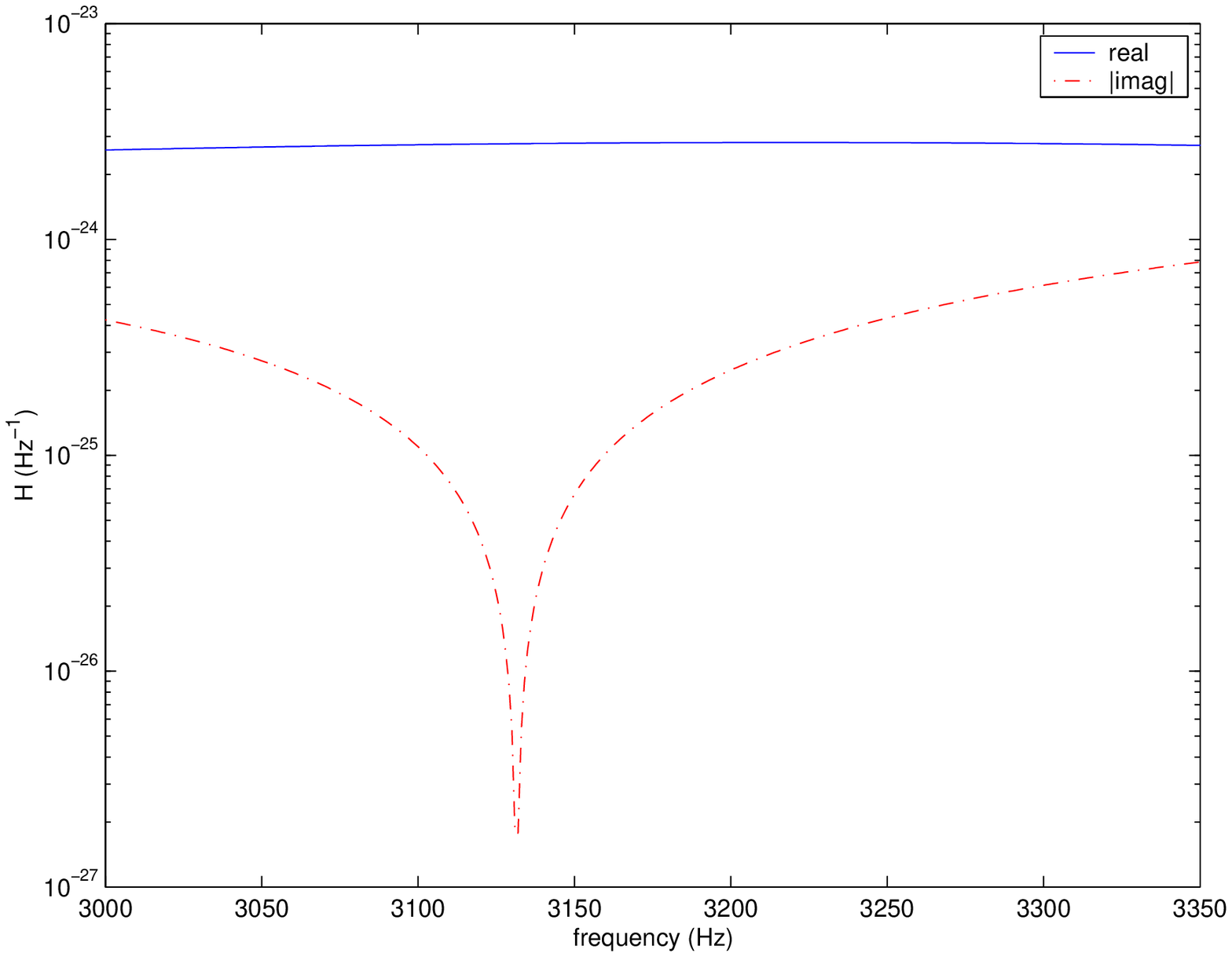, width=5cm, angle=0}}  \\
\hbox{(a)}    &  \hbox{(b)}                  \\
\end{tabular}
\end{center}
\caption{Waveform of a black hole ringdown (a) in the time domain
and (b) in the frequency domain.}
\end{figure}
\par Figure \ref{fig:waveform}.a shows the waveform $h(t)$ in the time
domain. The Fourier transform of the impulse $h(t)$ is defined as
\begin{equation}\label{eq:Hw}
H(\omega)=\int_{-\infty}^\infty h(t)e^{-i\omega t}dt.
\end{equation}
It represents the signal in the frequency domain and is shown in
Figure \ref{fig:waveform}.b. This figure shows that the signal
spectrum on SCHENBERG bandwidth ($3.17-3.24\mathrm{kHz}$) is
almost plane. This is a direct consequence of the low value for
the BH's quality factor ($Q\approx 2.16$, for that simulated
signal). For higher values of the quality factor the damping
function tends to a sinusoid and a peak may appear in the
spectrum. This would facilitate the detection of the signal.

\section{The SCHENBERG model}\label{sec:model}

We modelled the SCHENBERG detector by assuming a linear elastic
theory. Adopting this approach we determined the mechanical
response of the system and obtained an expression for the case
when six 2-mode mechanical resonators are coupled to the antenna's
surface according to the arrangement suggested by Johnson and
Merkowitz, the truncated icosahedron configuration
\cite{johnson1993,merkowitz1997}. We found an analytic expression
to calculate the spectral density for the mode channels, given by
\cite{costa2003}
\begin{equation}\label{eq:gspecdens}
S^g_m(\omega)=|\xi_m(\omega)|^2S^{\tilde
F^S}_m+\sum_{j=1}^6|\Omega^{R_1}_{mj}(\omega)|^2S^{\tilde
F^N_{R_1}}_j+\sum_{j=1}^6|\Omega^{R_2}_{mj}(\omega)|^2S^{\tilde
F^N_{R_2}}_j.
\end{equation}
The scalar $\xi_m(\omega)$ represents the transfer function of the
sphere's internal forces ${\tilde F^S}_m$ at angular frequency
$\omega$ on mode $m$. The matrix elements
$\Omega^{R_i}_{mj}(\omega)$ correspond to the response functions
of channel $m$ to the noise forces $\tilde F^{N_{R_i}}_j$ that are
acting on the mode $i$ of resonator $j$. $S^X$ denotes the
spectral density of the quantity $X$. In order to find equation
\ref{eq:gspecdens}, we assumed that all noise sources and possible
signals were statistically independent so the cross terms could be
neglected.
\par The GW effective force \cite{costa2003,zhou1995,harry1996} that acts
on the sphere (in the frequency domain) has spectral density
\begin{equation}\label{eq:SFm}
S^{\tilde\mathcal{F}^{GW}}_m(\omega)=\left(\frac{1}{2}\omega^2m_S\chi
R \right)^2 S^{\tilde h}_m(\omega).
\end{equation}
The sensitivity curve for mode $m$ is given through
\begin{equation}\label{eq:hm}
\tilde{h}_m(\omega)=\sqrt{\frac{1}{\left(\frac{1}{2}\omega^2m_S\chi
R \right)^2}\frac{S^g_m(\omega)}{|\xi_m(\omega)|^2}}.
\end{equation}
\begin{figure}\label{fig:sens_1st_goal}
  \begin{center}
  \includegraphics[width=8.5cm]{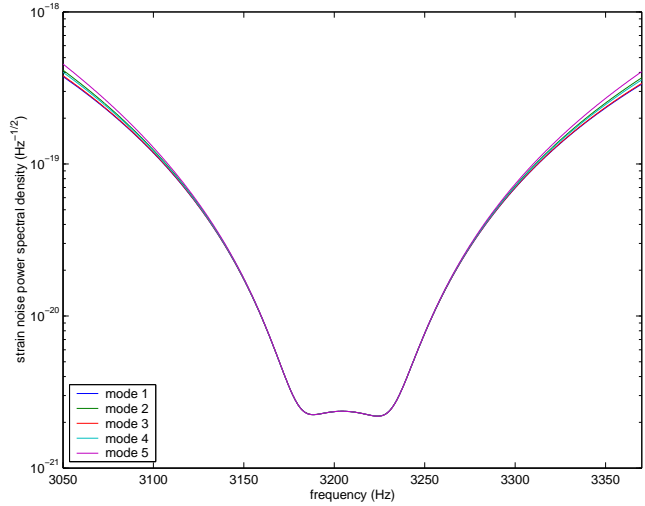}
  \caption{Sensitivity curve of SCHENBERG at $4.2\mathrm{K}$
  for the quadrupolar normal modes $m$.}
  \end{center}
\end{figure}
Figure \ref{fig:sens_1st_goal} shows the sensitivity curves for
the five quadrupolar normal modes of the sphere. We obtained this
curve assuming that no signal was present and that only the usual
noises in this kind of detector (Brownian noise, serial noise,
back-action noise and electronic phase and amplitude
noises)\footnote{For details on the parameters used in the
Schenberg model see Aguiar, O. D. \textit{et al}. and Frajuca, C.
\textit{et al}. in these Proccedings.} were disturbing the antenna
at $4.2\mathrm{K}$. The absence of a signal made filtering
unnecessary in this calculation. Also, we assumed a perfectly
symmetric sphere, which implies identical transfer functions
$\xi_m(\omega)$. Asymmetries on the sphere could make the transfer
function of a mode different from the transfer function of the
others.

\section{Signal-to-noise ratio}

Using the results in sections \ref{sec:signal} (which inform about
the signal) and \ref{sec:model} (which inform about the noise) it
is easy to calculate the integrated signal-to-noise ratio ($SNR$).
From the literature \cite{michelson1981} we have
\begin{equation}\label{eq:SNR}
SNR=\frac{1}{2\pi}\int_{-\infty}^{+\infty}\frac{|H(\omega)|^2}{Sn(\omega)}d\omega.
\end{equation}
The integral in equation \ref{eq:SNR}, which is dominated by the
interval of SCHENBERG bandwidth, gives $SNR\gtrsim 1$ for the
simulated signal when the BH's distance is $r\sim 20\mathrm{kpc}$.

\section{Conclusion}

If a low-spinning BH ($\hat{a}\sim 0.16$) with mass $\sim
4M_\odot$ and emission of about $1\%$ of its mass-energy in the
form of GW exists up to a distance of $\sim 20\mathrm{kpc}$, then
the SCHENBERG detector is expected to observe it when operating at
$4.2\mathrm{K}$.
\par In the calculation we didn't take into account other possible dependencies
on the BH parameters except the known frequency-mass dependence.
Strong dependencies between $\epsilon$ and $\hat{a}$ may make
signals from high-spinning BHs much stronger.
\par There are theories that claim the existence of BHs
but experimental evidences are indirect and inconclusive.
Certainly it is not yet possible to predict the correct number of
galactic BHs. However, the radius of $\sim 20\mathrm{kpc}$
includes the galactic center and the majority of the spiral arms,
which are the most probable birthplaces for stellar BHs. So the
detection of the kind of signal investigated in this work is an
opportunity for gravitational wave astronomers to make essential
discoveries about the black hole population in our galaxy.

\ack{This work was supported by \textbf{FAPESP} (under grant
numbers 1998/13468-9, 2001/14527-3 and 2003/02912-5),
\textbf{CNPq} (under grant number 300619/92-8) and
\textbf{MCT/INPE}.}


\section*{References}

\begin{thebibliography}{13}

\bibitem{echeverria1989}
F. Echeverria, Phys. Rev. D 40, 3194 (1989).

\bibitem{leaver1985} E. W. Leaver, Proc. R. Soc. London, Ser. A 402, 285 (1985).

\bibitem{creighton1999} J. D. E. Creighton, Phys. Rev. D 60, 022001-1
(1999).

\bibitem{GRASP} J. D. E. Creighton, GRASP 1.9.8 - Users Manual, ed. B. Allen, p. 251 (1999).

\bibitem{johnson1993} W. W. Johnson and S. M. Merkowitz, Phys. Rev.
Letters, 70, 2367 (1993).

\bibitem{merkowitz1997} S. M. Merkowitz and W. W. Johnson, Phys. Rev. D, 56, 7513 (1997).

\bibitem{costa2003} C. A. Costa, O. D. Aguiar and N. S. Magalh\~aes, (in preparation).

\bibitem{zhou1995} C. Z. Zhou, Phys. Rev. D, 51, 2517 (1995).

\bibitem{harry1996} G. M. Harry, T. R. Stevenson and H. J. Paik, Phys. Rev. D, 54, 2409 (1996).

\bibitem{michelson1981} P. F. Michelson and R. C. Taber, J. Appl. Phys., 52, 4313 (1981).

\end{thebibliography}
\end{document}